\documentclass[%
 reprint,
 amsmath,amssymb,
 aps,
]{revtex4-2}

\usepackage{graphicx}
\usepackage{dcolumn}
\usepackage{bm}

\usepackage{subfiles}
\usepackage{amsmath}
\usepackage{esvect}
\usepackage{braket}
\usepackage{autobreak}
\usepackage{comment}
\usepackage{here}
\usepackage{multirow}
\usepackage{color}
\usepackage{ulem}

\DeclareRobustCommand{\erase}{\bgroup\markoverwith{\textcolor{red}{\rule[.5ex]{2pt}{0.4pt}}}\ULon}
\DeclareRobustCommand{\eraseb}{\bgroup\markoverwith{\textcolor{blue}{\rule[.5ex]{2pt}{0.4pt}}}\ULon}

\begin{document}

\preprint{APS/123-QED}

\title{Excitation spectra of heavy baryons in a quark-diquark model
with relativistic corrections}

\author{S. Kinutani$^1$}
\author{H. Nagahiro$^{1,2}$}%
\author{D. Jido$^3$}
\affiliation{%
$^1$Department of Physics, Nara Women's University, Nara 630-856, Japan\\
$^2$Research Center for Nuclear Physics (RCNP), Osaka University, Ibaraki, Osaka 567-0047, Japan\\
$^3$Department of Physics, Tokyo Institute of Technology, 2-12-1 Ookayama, Megro, Tokyo 152-8551, Japan
}

\date{\today}

\begin{abstract}
    The excitation spectra of $\Lambda_c$ and $\Lambda_b$ baryons
    are investigated by using a quark-diquark model in which a single-heavy baryon
    is treated as the bound state of a heavy quark and a scalar diquark.
    We take two types of relativistic corrections into account
    for the quark-diquark potential.
    In the first type, we consider the one-gluon exchange between
    the heavy quark and one of the light quarks
    in the diquark.
    In the second, we consider the one-gluon exchange between a scalar particle
    and a heavy quark.
    We find that there is a large difference between two types of corrections
    due to different treatment of
    the internal color structure of the diquark.
    The relativistic corrections are important for the solution
    to the string tension puzzle, particularly, the Darwin term makes a large contribution.
\end{abstract}

\maketitle


    \section{Introduction}
    Understanding the structure of hadrons is one of the most important
    topics in hadron physics.
    It is too complex to describe the properties of hadrons in
    quantum chromodynamics (QCD) directly,
    since QCD is nonperturbative at low energies,
    so identifying the effective degree of freedom for hadrons is essential.
    The constituent quark plays a key role as the effective degree of freedom inside hadron,
    and, a diquark which is a two-quark pair correlation may also be
    one~\cite{Ida:1966ev,Lichtenberg:1967zz,Anselmino:1992vg,Barabanov:2020jvn}.
    This is particularly true of a diquark with an antisymmetric color, flavor, and spin,
    also known as a good diquark,
    as it has the most attractive correlation~\cite{Jaffe:2005zz}.
    The structure of hadrons are investigated in the viewpoint of diquark,
    for example, for baryons~\cite{Ida:1966ev,Lichtenberg:1967zz,Goldstein:1979wba,Lichtenberg:1969sxc,Lichtenberg:1982jp,Hernandez:2008ej,Lee:2009rt,Jido:2016yuv,Eichmann:2016yit,Kumakawa:2017ffl,Amano:2019jek,Amano:2021spn,Nielsen:2018uyn,Kim:2011ut,Liu:1983us} and light scalars
    \cite{Jaffe:1976ig,Black:1999dx,Maiani:2004uc,tHooft:2008rus}.

    Single-heavy baryons (called heavy baryons hereafter) particularly show promise for
    reaching the properties of diquarks
    due to the mass difference between the light quarks and the heavy quark
    \cite{Hernandez:2008ej,Lee:2009rt,Jido:2016yuv,Eichmann:2016yit,Kumakawa:2017ffl,Amano:2019jek,Amano:2021spn,Kim:2011ut}.
    The $\Lambda_c(2595)$ and $\Lambda_c(2625)$ baryons
    are
    the lowest excited states
    and can be interpreted as excitation of the $\lambda$-mode
    which is the relative motion between the heavy quark and the center of mass of light quarks,
    because these baryon can be regarded as spin-orbit (LS) partners of the rotational excitation
    of the heavy quark~\cite{Yoshida:2015tia}.
    With this speculation the light quark component can be considered as a good diquark.
    The important point here is that the diquark correlation works
    as an effective degree of freedom in the heavy baryon.

    The mass spectra of heavy baryons have been investigated in quark-diquark models
    in many previous works.
    In Ref.~\cite{Kim:2011ut}, the masses of the ground state of $\Lambda$, $\Lambda_c$, and $\Lambda_b$
    baryons were calculated in the QCD sum rule in which the diquark was introduced as an elementary field.
    They estimates the constituent diquark mass to 0.4 GeV and
    found that the QCD sum rule works well for these baryons.
    In Ref.~\cite{Jido:2016yuv}, the excitation spectra of $\Lambda_c$ and $\Lambda_b$ baryons were
    calculated in a quark-diquark model with the Coulomb-plus-linear-type potential.
    The diquark there is assumed to be a point-like good diquark,
    and the heavy baryons
    are treated as the bound systems of the heavy quark and diquark.
    The confinement potential may depend only on the color charge.
    If the diquark in the heavy baryon has the same color as the antiquark,
    it would be reasonable to use the same potential for both quark-antiquark and
    quark-diquark systems.
    However, Ref.~\cite{Jido:2016yuv} found the possibility
    that the confinement force between the quark and the diquark
    was about half of that
    in the quark-antiquark system in order to reproduce the experimental value of
    the $\Lambda_c$ $1p$ excitation energy.
    In Ref.~\cite{Kumakawa:2017ffl}, this puzzle,
    which we call a string tension puzzle hereafter,
    was tackled by considering
    the diquark size for the calculation of
    the excitation spectra of the $\Lambda_c$ baryon.
    According to this work, the size effect of the diquark reduces the excitation energy,
    and the diquark size
    $\rho \simeq 1.1$ fm reproduces the $\Lambda_c$ $1p$ excitation energy of 0.33 GeV.
    However, the spin-dependent force is not included in the potential.

    In this paper,
    we consider relativistic corrections
    to the quark-diquark potential as an alternative approach to this puzzle.
    Assuming that one heavy quark and light scalar diquark compose
    the heavy baryon,
    we consider two types of relativistic corrections
    for the quark-diquark potential.
    The diquark inside the heavy baryon is assumed
    to have
    anti-color $\bar{\bf 3}$ and spin singlet $S=0$.
    One is derived by considering a one-gluon exchange between
    the heavy quark and the light quarks in the diquark
    and the other by considering a one-gluon exchange between the quark and
    the scalar diquark.
    In the former approach,
    the relativistic corrections are calculated so that
    the diquark is composed of two light quarks,
    and in the latter,
    the diquark is assumed to be a scalar particle
    not having an internal color structure.
    Since the confinement force mainly depends on only the color charge and
    the good diquark has the same color charge as the antiquark in quarkonia,
    we may determine the parameters of the potential in the quark-diquark system
    by quarkonium spectra.
    We will see that the Darwin term plays an important role
    to improve the model calculation.
    We will also calculate the $\Xi_c$ excitation spectra to check the consistency of this model.

    This article is organized as follows.
    In Sec.~\ref{sec:Formulation},
    we calculate two types of the one-gluon exchange potential
    to introduce the relativistic corrections.
    In Sec.~\ref{sec:NumericalResults},
    we determine model parameters to reproduce
    the mass spectra of the charmonium,
    and show the excitation spectra of the $\Lambda_c$, $\Lambda_b$, and $\Xi_c$ baryons
    calculated with those parameters.
    Section~\ref{sec:summary} is devoted to the summary.


    \section{Formulation}
    \label{sec:Formulation}
    We describe the heavy baryon in a quark-diquark model.
    The diquark is assumed as the point-like good diquark which has antisymmetric
    color, flavor, and spin under the exchange of quarks in the diquark.

    We introduce two types of relativistic corrections whose
    differences come from the consideration of the color structure of the scalar diquark.
    One is that the color structure of the scalar diquark is considered
    by treating
    the diquark as a pair of two fermions.
    The relativistic corrections are derived from the interaction between the heavy quark and one of
    the light-quarks inside the scalar diquark.
    We call this potential $q$-$Q$ type potential.
    The other is that the internal color structure of the scalar diquark is not considered and
    the diquark is just treated as a scalar particle with color $\bar{\textbf{3}}$.
    The relativistic corrections are derived from the interaction between the scalar particle and
    the heavy quark.
    We refer this potential $S$-$Q$ type potential.

    We calculate the matrix elements of the considering quark-diquark potential $V$
    for each partial wave to obtain the effective potential $V_{\rm eff}$ as
    \begin{equation}
        V_{\rm eff}(r)= \braket{^{2S+1}L_J|V|^{2S+1}L_J}
    \end{equation}
    which is calculated below in the following subsections.
    We write the angular momentum state as $\ket{^{2S+1}L_J}$ with
    the total spin $S$ and the total angular momentum $J$.
    Then, we can get the Schr\"{o}dinger equation describing this system as
    \begin{equation}
        \left[
        -\frac{1}{2\mu}\frac{1}{r}\frac{d^2}{dr^2}r + 
            V_{\rm eff}(r) +V_0 + \frac{L(L+1)}{2\mu r^2}
        \right] R(r)
        =ER(r).
        \label{eq:schrodinger}
    \end{equation}
    Here $R(r)$ is the radial wave function, $\mu=\frac{m_d m_Q}{m_d+m_Q}$ is
    the reduced mass with the diquark mass $m_d$
    and the heavy quark mass $m_Q$, and $V_0$ is a constant.
    The total energy of the system is given by $E$.
    Since we are interested in the excitation spectra of heavy baryons,
    the constant $V_0$ is irrelevant to this analysis.
    
    \subsection{$q$-$Q$ type potential}
    \label{sec:qq_potential}
    \begin{figure*}
        \centering
        \includegraphics[scale=0.5]{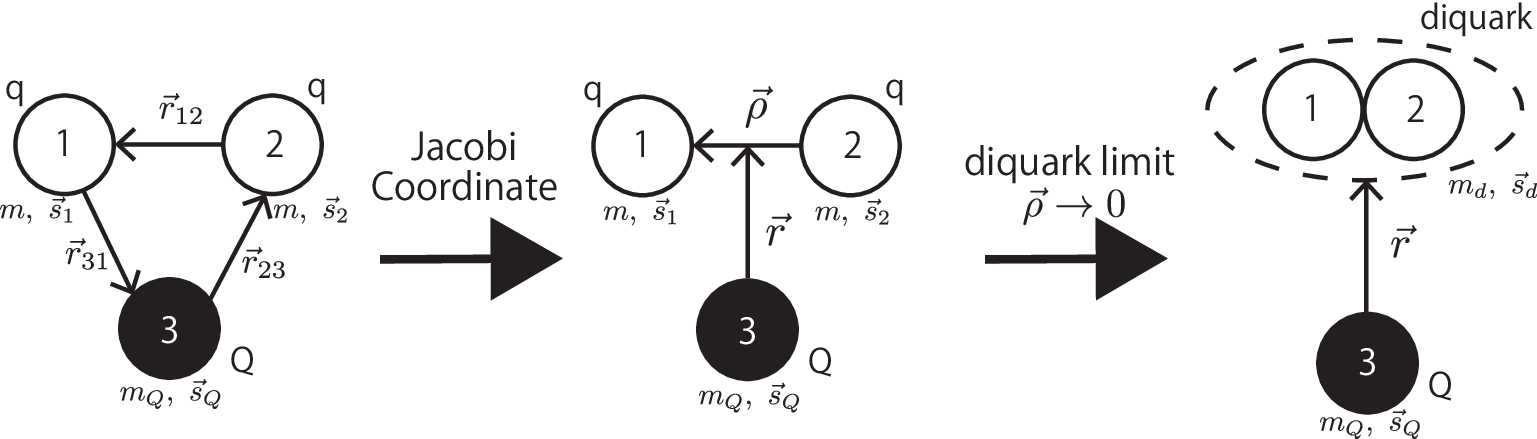}
        \caption{Schematic diagrams of the heavy baryon in a quark-diquark model with the Jacobi coordinate.
        Quark-1 and quark-2 denote light quarks having masses $m$, and quark-3 denotes the heavy quark having mass $m_Q$.
        By taking the diquark limit $\vec{\rho}\rightarrow0$, we treat the heavy baryon as the bound system of the diquark and heavy quark.}
        \label{fig:Jacobi_diquark_limit}
    \end{figure*}
    In this subsection, we derive $q$-$Q$ type quark-diquark potential
    with the QCD Breit-Fermi potential.
    We construct the two-body potential for the quark-diquark system
    by summing the quark-quark interactions between one of the light quarks and the heavy quark,
    and taking the distance between the two light quarks
    to zero $(\rho \rightarrow 0)$ as in Fig.~\ref{fig:Jacobi_diquark_limit},
    for the point-like diquark.

    We start from the quark-quark potential.
    The scattering amplitude $M_{qq}$ with the one-gluon exchange
    between quark-1 and quark-2
    is expressed as
    \begin{eqnarray}
        -iM_{qq}&=& 4 \pi \frac{2}{3} \alpha_s
        \bar{u}(\vec{p}^{\,\prime}_1)\gamma^{\mu}u(\vec{p}_1)D_{\mu \nu}(\vec{q})
        \bar{u}(\vec{p}^{\,\prime}_2)\gamma^{\nu}u(\vec{p}_2),
        \label{eq:amplitude_qq}
    \end{eqnarray}
    where $\alpha_s$ is the strong fine structure constant 
    and $D_{\mu \nu}$ is the gluon propagator.
    Here we consider a quark pair with color $\bar{\textbf{3}}$ and factor $2/3$ in Eq.~(\ref{eq:amplitude_qq})
    is a color factor for the $\bar{\textbf{3}}$ quark pair.
    The gluon three-momentum $\vec{q}$ is given by
    $\vec{q}=\vec{p}^{\,\prime}_1-\vec{p}_1=\vec{p}_2-\vec{p}^{\,\prime}_2$
    with momenta $\vec{p}_1$ and $\vec{p}^{\,\prime}_1$ of quark-1 in the initial and final states
    and ones $\vec{p}_2$ and $\vec{p}^{\,\prime}_2$ of quark-2 in the initial and final states.
    The quark spinor is given as
    \begin{equation}
        u(\vec{p})=\sqrt{\frac{E+m}{2E}}\left(
            \begin{array}{c}
                \chi \\
                \frac{\vec{\sigma}\cdot\vec{p}}{E+m}\chi
            \end{array}
        \right),
        \label{eq:spinor}
    \end{equation}
    where $\chi$ denotes a two component spinor and
    $\vec{\sigma}$ is the Pauli matrix.
    Here, the normalization factor is determined to satisfy
    the normalization $u^{\dagger}u=1$,
    because the wave function in the Schr\"{o}dinger equation is
    normalized in such a way~\cite{BA03038604,donoghue2014dynamics,DeRujula:1975qlm}.
    
    We reduce the amplitude of Eq.~(\ref{eq:amplitude_qq}) in the small
    momentum expansion up to $O(\frac{\vec{p}^{\,2}}{m^2})$ as
    \begin{eqnarray}
        M_{qq}
        &=&4\pi \frac{2\alpha_s}{3}\frac{-1}{\vec{q}^{\,2}}
        \chi'^{\dagger}_1\chi^{\dagger}_1\chi'^{\dagger}_2\chi^{\dagger}_2
        \nonumber\\
        &&\times \left[
            1-\left( \frac{1}{8m_1^2}+\frac{1}{8m_2^2} \right)\vec{q}^{\,2}\right. \nonumber \\
            &&~~\left.+\frac{i\vec{\sigma}_1\cdot(\vec{q}\times\vec{p}_1)}{4m_1^2}
            -\frac{i\vec{\sigma}_2\cdot(\vec{q}\times\vec{p}_2)}{4m_2^2} \right. \nonumber \\
            &&~~\left. - \frac{\vec{q}^{\,2}}{4m_1m_2}\vec{\sigma}_1\cdot\vec{\sigma}_2
            +\frac{(\vec{\sigma}_1\cdot\vec{q})(\vec{\sigma}_2\cdot\vec{q})}{4m_1m_2}
            \right. \nonumber \\
            &&\left.-\frac{i\vec{\sigma}_1\cdot(\vec{q}\times\vec{p}_2)}{2m_1m_2}
            +\frac{i\vec{\sigma}_2\cdot(\vec{q}\times\vec{p}_1)}{2m_1m_2}\right. \nonumber \\
            &&\left.-\frac{1}{m_1m_2}\left(\vec{p}_1\cdot\vec{p}_2-\frac{(\vec{p}_1\cdot\vec{q})(\vec{p}_2\cdot\vec{q})}{\vec{q}^{\,2}}\right)
        \right]\nonumber \\
        &&\chi'_1\chi_1\chi'_2\chi_2,
        \label{eq:amp_detail}
    \end{eqnarray}
    where $m_1$ and $m_2$ are masses of quark-1 and quark-2, respectively.
    
    The quark-quark potential $V_{qq}$ is defined by the Fourier transform
    of the scattering amplitude (\ref{eq:amp_detail}) as
    \begin{eqnarray}
        V_{qq}(\vec{p}_1,\vec{p}_2,\vec{r})=\int{
            \frac{d^3q}{(2\pi)^3}e^{-i\vec{r}\cdot\vec{q}}M_{qq}(\vec{p}_1,\vec{p}_2,\vec{q})
        },
        \label{eq:int_Vqq}
    \end{eqnarray}
    where $\vec{r}$ is the relative coordinate between two quarks.
    Thus we obtain the quark-quark potential as
    \begin{eqnarray}
        &&V_{qq}(\vec{p}_1,\vec{p}_2,\vec{r}) \nonumber \\
        &=&-\frac{2}{3}\alpha_s\left\{
            \frac{1}{r}
             - \frac{8\pi}{3m_1m_2}\vec{s}_1\cdot\vec{s}_2 \delta^3(\vec{r}) \right. \nonumber \\
            &&\left.-\frac{\pi}{2}\left(\frac{1}{m_1^2}+\frac{1}{m_2^2} \right)\delta^3(\vec{r}) \right. \nonumber \\
            &&\left.
            -\frac{1}{m_1m_2}\frac{1}{r^3}\left[ \frac{3(\vec{s}_1\cdot\vec{r})(\vec{s}_2\cdot\vec{r})}{r^2}
            -\vec{s}_1\cdot\vec{s}_2 \right] \right. \nonumber \\
            &&\left.
            +\frac{1}{r^3}\left[ \frac{\vec{s}_1\cdot(\vec{r}\times\vec{p}_1)}{2m_1^2}
            -\frac{\vec{s}_2\cdot(\vec{r}\times\vec{p}_2)}{2m_2^2} \right.\right. \nonumber \\
            &&\left.\left.
            +\frac{\vec{s}_1\cdot(\vec{r}\times\vec{p}_2)-\vec{s}_2\cdot(\vec{r}\times\vec{p}_1)}{m_1m_2}
            \right]
            \right. \nonumber \\
            &&\left.
                -\frac{1}{2m_1m_2}\frac{1}{r}\left(
                    \vec{p}_1\cdot\vec{p}_2+\frac{(\vec{r}\cdot\vec{p}_1)(\vec{r}\cdot\vec{p}_2)}{r^2}
                \right)
        \right\}.
        \label{eq:qq_qqbar_potential}
    \end{eqnarray}
    Summing all interactions in the considering system,
    we obtain so-called QCD Breit-Fermi potential.

    Now, we derive the two-body potential for the quark-diquark system
    from the interactions between the heavy quark and the light quarks.
    The spacial coordinate of the center of mass $\vec{R}$ is written as
    \begin{equation}
        \vec{R} = \frac{m(\vec{r}_1+\vec{r}_2) +m_Q\vec{r}_3 }{2m+m_Q}.
    \end{equation}
    Here, $m$ is the light quark mass
    and $m_Q$ is the heavy quark mass.
    The relative coordinate $\vec{\rho}$ between light quarks and
    the relative coordinate $\vec{r}$ between a heavy quark and the center of mass of
    light quarks are written as
    \begin{eqnarray}
        \vec{\rho}&=&\vec{r}_1-\vec{r}_2,\\
        \vec{r}&=&\frac{1}{2}(\vec{r}_1+\vec{r}_2)-\vec{r}_3.
    \end{eqnarray}
    After considering
    the interaction between the light quark and the heavy quark,
    we take the diquark limit $\vec{\rho}\rightarrow 0$.

    The total spin $\vec{S}$ of the system is obtained by sum of
    the heavy quark spin $\vec{s}_Q$ and the diquark spin $\vec{s}_d$ as
    \begin{equation}
        \vec{S}= \vec{s}_d + \vec{s}_Q,
    \end{equation}
    where diquark spin is given by the sum of spins of
    two light quarks $\vec{s}_1$ and $\vec{s}_2$ as
    \begin{eqnarray}
        \vec{s}_d &=&\vec{s}_1 +\vec{s}_2.
        \label{eq:diquark_spin}
    \end{eqnarray}
    Since we consider a good diquark, the diquark spin is $\vec{s}_d = 0$.
    The total angular momentum $\vec{J}$ is written as
    \begin{equation}
        \vec{J}=\vec{L}+\vec{S},
    \end{equation}
    where the relative orbital angular momentum between a heavy quark and a diquark
    is $\vec{L}$.
    We can get the quark-diquark potential
    by the sum of the confinement part $V_{\rm conf}$ and the non-confinement part obtained
    from the QCD Breit-Ferim potential.
    In the following, we calculate each term of the effective $q$-$Q$ type potential $V^{(qQ)}_{Qd}$ separately.
    Here we allow to take an individual coupling constant for each term.

    We first calculate the confinement term of the total potential.
    We consider a liner type confinement potential for a quark pair,
        which is given as $\frac{1}{2}k r$ with the string tension $k$ and
        the distance between two quarks.
        The string tension is determined by the quark-antiquark system and factor $1/2$
        is the relative color factor between the quark-quark and quark-antiquark systems.
        The confinement term is written as
        \begin{eqnarray}
            \frac{1}{2}k\left\{
                |\vec{r}_{23}|+|\vec{r}_{31}|
            \right\}
            = k|\vec{r}|
        \end{eqnarray}
        by taking the diquark limit $\vec{\rho} \rightarrow 0$,
        that is, setting the distance between two light quarks to zero.
        Thus, the confinement potential $V_{\rm conf}$ is 
        \begin{eqnarray}
            V_{\rm conf} = kr.
        \end{eqnarray}

    The Coulomb term is written as
        \begin{eqnarray}
            &&-\frac{2}{3}\alpha_{\rm Coul}\left\{
                \frac{1}{|\vec{r}_{23}|} + \frac{1}{|\vec{r}_{31}|}
            \right\}
            \nonumber \\
            &&= -\frac{4}{3} \frac{\alpha_{\rm Coul}}{|\vec{r}|}
        \end{eqnarray}
        after taking the diquark limit $\vec{\rho} \rightarrow 0$.
        The effective Coulomb term $V_{\rm Coul}$
        can be calculated as
        \begin{eqnarray}
            V_{\rm Coul}
            = -\frac{4}{3}\frac{\alpha_{\rm Coul}}{r}.
            \label{eq:effterm_Coul}
        \end{eqnarray}

        We write the Darwin term for the effective $q$-$Q$ potential
        by taking the diquark limit $\vec{\rho} \rightarrow 0$
        as
        \begin{eqnarray}
            &&\frac{2}{3}\alpha_{\rm Dar}\frac{\pi}{2}
            \left\{
                \delta^3(\vec{r}_{23})\left( \frac{1}{m^2}  + \frac{1}{m_Q^2} \right)
            \right. \nonumber \\
            &&~~~~~~~~~~~~~\left. +\delta^3(\vec{r}_{31})\left( \frac{1}{m_Q^2} + \frac{1}{m^2} \right)
            \right\}
            \nonumber \\
            &&= \frac{2\pi}{3}\alpha_{\rm Dar}\delta^3(\vec{r})
            \left( \frac{4}{m_d^2} + \frac{1}{m_Q^2}\right),
        \end{eqnarray}
        where we have assumed that the light quark mass is given by a half of
        the diquark mass $m_d$.
        The expectation value of the delta function can be
        evaluated by the value of the wave function at the origin
        as
        \begin{eqnarray}
            \int d^3r\psi^*(\vec{r})\delta^3(\vec{r})\psi(\vec{r})
            = \frac{2L+1}{4\pi} |R_{nL}(0)|^2.
            \label{eq:expectation_delta}
        \end{eqnarray}
        Thus, we obtain the effective Darwin term $V_{\rm Dar}$ as
        \begin{eqnarray}
            V_{\rm Dar} = \frac{2\pi}{3}\alpha_{\rm Dar}\left(
                \frac{1}{m_Q^2}+\frac{4}{m_d^2}
            \right)
            \frac{2L+1}{4\pi r^2}\delta(r).
            \label{eq:effterm_Dar}
        \end{eqnarray}
        Later we will regularize the delta function to reduce the singularity at the origin,
        and thus the Darwin term can contribute also to higher partial waves
        but its contributions are highly suppressed by the centrifugal barrier.

        The hyperfine interaction is
        written as
        \begin{eqnarray}
            &&\frac{1}{2}\frac{32\pi}{9}\frac{\alpha_{\rm Hyp}}{mm_Q}
            \left\{
                \delta^3(\vec{r}_{23})\vec{s}_2\cdot\vec{s}_Q
                +\delta^3(\vec{r}_{31})\vec{s}_Q\cdot\vec{s}_1
            \right\}
            \nonumber \\
            &&= \frac{32\pi}{9}\alpha_{\rm Hyp}
            \frac{1}{m_d m_Q}\delta^3(\vec{r})
                \vec{s}_d\cdot\vec{s}_Q.
        \end{eqnarray}
        by taking the diquark limit $\vec{\rho}\rightarrow0$.
        Since the good diquark has the spin $\vec{s}_d=0$,
        the effective hyperfine interaction term $V_{\rm Hyp}$
        for the scalar diquark is
        \begin{eqnarray}
            V_{\rm Hyp} = 0.
            \label{eq:effterm_Hyp}
        \end{eqnarray}

        The spin-orbit interaction of the $q$-$Q$ type potential is
        expressed as
        \begin{eqnarray}
            &&\frac{1}{2}\frac{4}{3}\alpha_{\rm LS} \nonumber \\
            &&\times\left\{
                \frac{1}{|\vec{r}_{23}|^3}\left[
                \left( \frac{1}{4m^2} + \frac{1}{4m_Q^2} +\frac{1}{mm_Q} \right)
                \vec{L}_{2Q}\cdot(\vec{s}_2+\vec{s}_Q) \right.\right. \nonumber \\
                &&\left.\left. +
                \left( \frac{1}{4m^2} - \frac{1}{4m_Q^2} \right)
                \vec{L}_{2Q}\cdot(\vec{s}_2-\vec{s}_Q)
            \right] \right. \nonumber \\
            &&\left.+
            \frac{1}{|\vec{r}_{31}|^3}\left[
                \left( \frac{1}{4m_Q^2} + \frac{1}{4m^2} +\frac{1}{mm_Q} \right)
                \vec{L}_{Q1}\cdot(\vec{s}_Q+\vec{s}_1) \right.\right. \nonumber \\
                &&\left.\left. +
                \left( \frac{1}{4m_Q^2} - \frac{1}{4m^2} \right)
                \vec{L}_{Q1}\cdot(\vec{s}_Q-\vec{s}_1)
            \right]
            \right\}\nonumber \\  
            &&= \frac{4}{3}\frac{\alpha_{\rm LS}}{|\vec{r}|^3}\frac{m_d+m_Q}{m_d+2m_Q}
                \left(
                    \frac{1}{2m_Q^2}+\frac{4}{m_d m_Q}
                \right)\vec{L}\cdot\vec{s}_Q
            \label{eq:LS_qq}    
        \end{eqnarray}
        by taking the diquark limit $\vec{\rho}\rightarrow0$,
        where the relative orbital angular momenta between quark-1 and the heavy quark
        and quark-2 and the heavy quark are written as $\vec{L}_{Q1}$ and $\vec{L}_{2Q}$,
        respectively.
        In the case of the point-like diquark $\vec{\rho} \rightarrow 0$,
        they can be written as
        \[\vec{L}_{Q1}=\vec{L}_{2Q}=\frac{m_d+m_Q}{m_d+2m_Q}\vec{L}\]
        with the relative orbital angular momentum $\vec{L}$ between the diquark and the heavy quark.
        The good diuqark spin $\vec{s}_d= \vec{s}_1+\vec{s}_2$ is zero.
        Thus integrating over a solid angle, we obtain the LS term $V_{\rm LS}$
        of the effective potential as
        \begin{eqnarray}
            V_{\rm LS}&=& \frac{4}{3}\frac{\alpha_{\rm LS}}{r^3}\frac{m_d+m_Q}{m_d+2m_Q}
            \left(
                \frac{1}{2m_Q^2}+\frac{2}{m_d m_Q}
            \right) \nonumber \\
             &&\times \frac{1}{2}\left(
                J(J+1)-L(L+1) - \frac{3}{4}
            \right).
            \label{eq:effterm_LS}
        \end{eqnarray}

        The tensor term is obtained as
        \begin{eqnarray}
            &&\frac{1}{2}\left\{
                \frac{4}{3mm_Q}\frac{\alpha_{\rm tens}}{|\vec{r}_{23}|}
                \left(
                    \frac{3(\vec{s}_2\cdot\vec{r}_{23})(\vec{s}_Q\cdot\vec{r}_{23})}{|\vec{r}_{23}|^2}
                    -\vec{s}_2\cdot\vec{s}_Q
                \right)\right. \nonumber \\
                && \left.
                +\frac{4}{3mm_Q}\frac{\alpha_{\rm tens}}{|\vec{r}_{31}|}
                \left(
                    \frac{3(\vec{s}_Q\cdot\vec{r}_{31})(\vec{s}_1\cdot\vec{r}_{31})}{|\vec{r}_{31}|^2}
                    -\vec{s}_Q\cdot\vec{s}_1
                \right)
            \right\}
            \nonumber\\
            &&=
            \frac{4}{3}\frac{\alpha_{\rm tens}}{|\vec{r}|^3}
            \frac{1}{m_dm_Q}
            \left(
                \frac{3(\vec{s}_d\cdot\vec{r})(\vec{s}_Q\cdot\vec{r})}{|\vec{r}|^2}
                +\vec{s}_d\cdot\vec{s}_Q
            \right)
        \end{eqnarray}
        by taking the diquark limit $\vec{\rho}\rightarrow 0$ and replacing the quark mass to the diquak mass.
        Since the diquark spin is $\vec{s}_d=0$,
        the part of the braket in the final line venishes.
        Therefore the effective tensor term $V_{\rm tens}$ is
        \begin{equation}
            V_{\rm tens} = 0.
            \label{eq:tensor}
        \end{equation}

        The orbit-orbit interaction on the $q$-$Q$ type potential is
        expressed as
        \begin{eqnarray}
            &&\frac{1}{2} \left\{
                \frac{2}{3mm_Q} \frac{\alpha_{\rm oo}}{|\vec{r}_{23}|}
            \left(
                \vec{p}_2\cdot\vec{p}_Q +\frac{(\vec{p}_2\cdot\vec{r}_{23})(\vec{p}_Q\cdot\vec{r}_{23})}{|\vec{r}_{23}|^2}
            \right)\right. \nonumber \\
            &&\left.+\frac{2}{3mm_Q} \frac{\alpha_{\rm oo}}{|\vec{r}_{31}|}
            \left(
                \vec{p}_Q\cdot\vec{p}_1 +\frac{(\vec{p}_Q\cdot\vec{r}_{31})(\vec{p}_1\cdot\vec{r}_{31})}{|\vec{r}_{31}|^2}
            \right)
            \right\}
            \nonumber\\
            &&=
            \frac{2}{3}\frac{1}{m_dm_Q}\frac{\alpha_{\rm oo}}{|\vec{r}|}\nonumber \\
            &&\times\left[
                    (\vec{p}_1+\vec{p}_2)\cdot\vec{p}_Q
                    + \frac{((\vec{p}_1+\vec{p}_2)\cdot\vec{r})(\vec{p}_Q\cdot\vec{r})}{|\vec{r}|^2}
            \right]
        \end{eqnarray}
        by taking the diquark limit $\vec{\rho}\rightarrow 0$ for the above expression
        and replacing the mass of the light quark
        to one of the diquark.
        The momentum of each quark can be rewritten with the total momentum $\vec{P}$
        and the relative momentum $\vec{p}$ as
        \begin{equation*}
            \vec{p}_1+\vec{p}_2 = \frac{m_d}{m_d+m_Q}\vec{P}+\vec{p}
        \end{equation*}
        and
        \begin{equation*}
            \vec{p}_Q= \frac{m_Q}{m_d+m_Q}\vec{P}-\vec{p}.
        \end{equation*}
        Thus, the orbit-orbit interaction $V_{\rm oo}$ is obtained as
        \begin{eqnarray}
           &&V_{\rm oo}
           =
            -\frac{2}{3}\frac{1}{m_dm_Q}\frac{\alpha_{\rm oo}}{r}
            \left(
                -\frac{2}{r}\frac{d^2}{dr^2}r + \frac{L(L+1)}{r^2}
            \right).
            \label{eq:effterm_oo}
        \end{eqnarray}

        Collecting the above results,
        we have the quark-diquark potential derived by the QCD Breit-Fermi potential as
        \begin{eqnarray}
            V^{(qQ)}_{Qd} &=& V_{\rm conf}+V_{\rm Coul}+V_{\rm Dar} \\ \nonumber
            &+&V_{\rm Hyp}+V_{\rm LS} +V_{\rm tens} +V_{\rm oo}.
        \end{eqnarray}
        We regularize the singularities at the origin and
        obtain the $q$-$Q$ type effective potential $V^{(qQ)}_{Qd}$ as
        \begin{widetext}
            \begin{eqnarray}
        V^{(qQ)}_{Qd}(r) &=& kr -\frac{4}{3}\frac{\alpha_{\rm Coul}}{r}
        +\frac{2\pi}{3}\alpha_{\rm Dar}\left(
                \frac{1}{m_Q^2}+\frac{4}{m_d^2}
            \right)
            \frac{2L+1}{4\pi r^2}\Lambda e^{-\Lambda^2 r^2} \nonumber \\
        &&+\frac{4}{3}\frac{\alpha_{\rm LS}}{r^3}\frac{m_d+m_Q}{m_d+2m_Q}
            \left(
                \frac{1}{2m_Q^2}+\frac{2}{m_d m_Q}
            \right)
            \frac{(1-e^{-\Lambda r})^2}{2}
            \left(
                J(J+1)-L(L+1) - \frac{3}{4}
            \right) \nonumber \\
        &&-\frac{2}{3}\frac{(1-e^{-\Lambda r})^2}{m_d m_Q}\frac{\alpha_{\rm oo}}{r}
            \left(
                -\frac{2}{r}\frac{d^2}{dr^2}r+ \frac{L(L+1)}{r^2}
            \right)
    \end{eqnarray}
        \end{widetext} 
    with the regularization parameter $\Lambda$.

    For later use, we also show the corresponding effective potential for the quark-antiquark $V_{Q\bar{Q}}$.
    According to the color factor the quark-antiquark interaction is
    twice as strong as the quark-quark potential.
    To this end, we take the factor $1$ instead of $1/2$ in Eq.~(\ref{eq:qq_qqbar_potential}).
    Each term of the regulalized effective quark-antiquark potential for the heavy quark with mass $m_Q$
    is expressed as
    \begin{equation}
        V_{Q\bar{Q}}= V_{\rm conf}+V_{\rm Coul} + V_{\rm Dar}
        + V_{\rm Hyp} + V_{\rm LS} +V_{\rm tens}
        + V_{\rm oo} 
    \end{equation}
    with
    \begin{equation}
        V_{\rm conf} = kr,
    \end{equation}
    \begin{equation}
        V_{\rm Coul} = -\frac{4}{3}\frac{\alpha_{\rm Coul}}{r},
    \end{equation}
    \begin{equation}
        V_{\rm Dar} = \frac{2\pi}{3}\alpha_{\rm Dar}\frac{2}{m_Q^2}
        \frac{2L+1}{4\pi r^2}\Lambda e^{-\Lambda^2 r^2},
    \end{equation}
    \begin{eqnarray}
        V_{\rm Hyp} &=& \frac{32\pi}{9m_Q^2}\alpha_{\rm Hyp}
        \frac{2L+1}{4\pi r^2}\Lambda e^{-\Lambda^2 r^2} \nonumber \\
        &&\times\frac{1}{2}\left(
            S(S+1)-\frac{3}{2}
        \right),
    \end{eqnarray}
    \begin{eqnarray}
        V_{\rm LS} &=& \frac{4}{3}\frac{\alpha_{\rm LS}}{r^3}\frac{3}{2m_Q^2}
        (1-e^{-\Lambda r})^2 \nonumber \\
        &&\times\frac{1}{2}(J(J+1)-L(L+1)-S(S+1)),
    \end{eqnarray}
    \begin{equation}
        V_{\rm tens} = \frac{4}{3m_Q^2}\frac{\alpha_{\rm tens}}{r^3}(1-e^{-\Lambda r})^2
        \alpha_{S,J},
        \label{eq:qqbar_Tensor}
    \end{equation}
    and
    \begin{eqnarray}
        V_{\rm oo} &=& -\frac{2}{3m_Q^2}\frac{\alpha_{\rm oo}}{r}\nonumber \\
        &&\times\left[
            -\frac{2}{r}\frac{d^2}{dr^2}r+\frac{L(L+1)}{r^2}
        \right](1-e^{-\Lambda r})^2,
    \end{eqnarray}
    where $\Lambda$ is the regulalization cut-off.
    The coefficient $\alpha_{S,J}$ appearing in Eq.~(\ref{eq:qqbar_Tensor}) is listed in
    \begin{equation}
        \begin{split}
            \alpha_{0,J}&=0,\\
            \alpha_{1,L-1}&=-\frac{L+1}{2L-1},\\
            \alpha_{1,L}&=1,\\
            \alpha_{1,L+1}&=-\frac{L}{2L+1}.
        \end{split}
        \label{eq:tens}
    \end{equation}
    The total spin $\vec{S}$ is the sum of heavy quark spins as $\vec{S}= \vec{s}_1+\vec{s}_2$, and
    the orbital angular momentum is written as $\vec{L}$, so
    the total angular momentum is $\vec{J}=\vec{S}+\vec{L}$.

    \subsection{$S$-$Q$ type potential}
    \label{sec:Sq_potential}
    In this subsection, we derive the quark-diquark potential
    by treating the diquark as the scalar particle with color $\bar{\textbf{3}}$.
    We consider the one-gluon exchange potential between the scalar particle and
    the remaining heavy quark~\cite{Goldstein:1979wba}.
    The color structure of the diquark is not considered in this model,
    which is the difference from the $q$-$Q$ type potential.

    We call the obtained quark-diquark potential $S$-$Q$ type potential.
    Since we assume that the scalar particle have
    anti-color $\bar{\bf 3}$,
    we take the strength of the coupling constant for the scalar-quark system
    to be the same as the quark-antiquark system.
    Here we use again the Breit equation \cite{BA03038604}
    to derive the effective interaction between the diquark and the heavy quark.

    The scattering amplitude $M_{SQ}$
    for the system of a scalar diquark and a heavy quark is expressed as
    \begin{eqnarray}
        -iM_{SQ}
        &=& 4\pi \frac{4}{3}\alpha_s\frac{(p_1'+p_1)^{\mu}}{\sqrt{2E_1'2E_1}}D_{\mu \nu}
        \bar{u}(\vec{p}^{\,\prime}_2)\gamma^{\nu}u(\vec{p}_2)
        \label{eq:Sq_amplitude}
    \end{eqnarray}
    with the momenta $\vec{p}_1$ and $\vec{p}^{\,\prime}_1$ of the scalar diquark in the initial and final states
    and those $\vec{p}_2$ and $\vec{p}^{\,\prime}_2$ of the heavy quark in the initial and final states.
    $E_1$ and $E_1'$ denote the energies of the scalar diquark in the initial and final states, respectively.
    Factor $4/3$ in Eq.~(\ref{eq:Sq_amplitude}) is the color factor of one gluon exchange
    for the quark and $\bar{\textbf{3}}$ diquark.
    The factor
    $\frac{1}{\sqrt{2E_1'2E_1}}$ is adapted as the normalization factor
    so that the time component of its current is unity.
    In fact, this causes the different expressions from
    the quark-diquark potential derived
    in Ref.~\cite{Goldstein:1979wba}.
    We discuss this difference and importance in detail after completing this derivation.

    We calculate the scattering amplitude in the non-relativistic expansion as
    \begin{eqnarray}
        M_{SQ} &=& 4\pi \frac{4\alpha_s}{3}\frac{-1}{\vec{q}^{\,2}}\chi'^{\dagger}
        \left[
            1-\frac{1}{8m_Q^2}\vec{q}^{\,2} \right. \nonumber \\
        &&\left.
            -\frac{i\vec{\sigma}\cdot(\vec{q}\times\vec{p}_2)}{4m_Q^2}
            +\frac{i \vec{\sigma}\cdot(\vec{q}\times \vec{p}_1)}{2m_d m_Q}
            \right. \nonumber \\
        &&\left. 
        -\frac{1}{m_d m_Q}\left( \vec{p}_1\cdot\vec{p}_2 - \frac{(\vec{p}_1\cdot\vec{q})(\vec{p}_2 \cdot \vec{q})}{\vec{q}^{\,2}} \right)
        \right]\chi.
        \label{eq:Sq_amp}
    \end{eqnarray}
    Performing Fourier transform of the scattering amplitude
    \begin{equation}
        V_{SQ}(\vec{p}_1,\vec{p}_2,\vec{r}) = \int{
            \frac{d^3 q}{(2\pi)^3} e^{-i \vec{r}\cdot\vec{q}}
            M_{SQ}(\vec{p}_1,\vec{p}_2,\vec{q})
        },
    \end{equation}
    we obtain the quark-diquark potential as
    \begin{eqnarray}
        V_{SQ}(\vec{p},\vec{r}) &=& -\frac{4}{3}\alpha_s\left[
        \frac{1}{r} -4\pi \frac{1}{8m_Q^2} \delta^3(\vec{r})
        \right. \nonumber \\
    &&\left.
    - \frac{1}{2m_dm_Qr}\left( \vec{p}^{\,2} +\frac{(\vec{p}\cdot\vec{r})^2}{r^2} \right)
    \right. \nonumber \\
    &&\left.
    -\frac{1}{r^3}\left(\frac{1}{2m_Q^2} + \frac{1}{m_dm_Q}\right)
    \vec{s}_Q\cdot\vec{L}
    \right]
    \end{eqnarray}
    with the relative momentum $\vec{p}=\vec{p}_1=-\vec{p}_2$,
    the spin of the heavy quark $\vec{s}_Q$, and
    the relative orbital angular momentum $\vec{L}$ between the scalar particle and the heavy quark.

    The $S$-$Q$ type quark-diquark potential is given by sum of
    the confinement term $V_{\rm conf}$ and the non-confinement terms obtained from
    one-gluon exchange between the scalar particle and quark as
    \begin{equation}
        V^{(SQ)}_{Qd}=
        V_{\rm conf} +V_{\rm Coul}+ V_{\rm Dar} + V_{\rm LS}+V_{\rm oo}.
    \end{equation}
    The difference from the $q$-$Q$ type potential is only the dependence of mass of the diquark.
    Thus
    the matrix elements can be obtained in the same way of the $q$-$Q$ type potential
    with different coefficients.
    We obtain the effective potential as follows.

    The effective confinement term $V_{\rm conf}$ can be obtained as
        \begin{equation}
            V_{\rm conf} = kr.
        \end{equation}
    The effective Coulomb term $V_{\rm Coul}$ is
        \begin{equation}
            V_{\rm Coul} = -\frac{4}{3}\frac{\alpha_{\rm Coul}}{r}.
        \end{equation}
    The effective Darwin term $V_{\rm Dar}$ is
        \begin{equation}
            V_{\rm Dar}=\frac{2\pi}{3}\alpha_{\rm Dar}\frac{1}{m^2_Q}
            \frac{2L+1}{4 \pi r^2}\delta(r).
            \label{eq:Derwin_Sq}
        \end{equation}
    The effective spin-orbit interaction $V_{\rm LS}$ is
        \begin{eqnarray}
            V_{\rm LS} &=& -\frac{4}{3}\frac{\alpha_{\rm LS}}{r^3}\left(
                \frac{1}{2m^2_Q}+\frac{1}{m_dm_Q}
            \right)\nonumber \\
            && \times\frac{1}{2}(J(J+1)-L(L+1)-s_Q(s_Q+1)).
        \end{eqnarray}
    The effective orbit-orbit interaction $V_{\rm oo}$ is
        \begin{equation}
            V_{\rm oo}=-\frac{4}{3}\frac{\alpha_{\rm oo}}{r}\frac{1}{2m_dm_Q}\left(
                -\frac{2}{r}\frac{d^2}{dr^2}r+\frac{L(L+1)}{r^2}
            \right).
            \label{eq:oo_Sq}
        \end{equation}

    Thus the regularized $S$-$Q$ type effective potential $V^{(SQ)}_{Qd}$ is
    \begin{widetext}
        \begin{eqnarray}
        {V^{(SQ)}_{Qd}(r)}
        &=&kr
        -\frac{4}{3}\frac{\alpha_{\rm Coul}}{r}
        +\frac{16\pi}{3}\frac{\alpha_{\rm Dar}}{8m_Q^2}
            \frac{2L+1}{4\pi r^2}\Lambda e^{-\Lambda^2 r^2}
                 \nonumber \\
            &&
                +\frac{4}{3}\frac{\alpha_{\rm LS}}{r^3} \left(
                    \frac{1}{2m_Q^2}+\frac{1}{m_dm_Q}
                \right)
                \frac{(1-e^{-\Lambda r})^2}{2}(J(J+1)-L(L+1)-s_Q(s_Q+1))\nonumber \\
            &&-\frac{4}{3}\frac{\alpha_{\rm oo}}{r}
                \frac{(1-e^{-\Lambda r})^2}{2m_dm_Q}\left( -\frac{2}{r}\frac{d^2}{dr^2}r +\frac{L(L+1)}{r^2}\right).
    \end{eqnarray}
    \end{widetext}
    The confinement term, Coulomb term, and the orbit-orbit interaction have the completely same forms
    as those on the $q$-$Q$ type potential.
    The spin-orbit interaction and
    the Darwin term have different coefficients
    from those of the $q$-$Q$ type potential.
    
    Now, we discuss the Darwin term.
    The Darwin term in the $S$-$Q$ type potential does not depends on the diquark mass $m_d$
    as in Eq.~(\ref{eq:Derwin_Sq}).
    This result for a scalar particle is consistent in Ref.~\cite{gross2008relativistic}.
    According to Ref.~\cite{gross2008relativistic}, the Darwin term appears from the order of $m^{-3}$
    for a scalar particle.
    On the contrary,
    the potential between the scalar particle and the quark in Ref.~\cite{Goldstein:1979wba}
    has the Darwin term containing the diquark mass in this order.
    Their scalar particle current
    is normalized so that the time component of its current is $2E$.
    Here we need to
    normalize the wave function of the scalar particle to satisfy that
    the time component of its current is unity
    because this is the way to normalize the wave function in Schr\"{o}dinger equation.
    The orbit-orbit interaction in our result also have different form from
    that in Ref.~\cite{Goldstein:1979wba} because of the difference of the normalization,
    and there are some missing terms.

    \section{Numerical Results}
    \label{sec:NumericalResults}
    In this section, we discuss our results of the excitation energy spectra of
    quarkonia and heavy baryons.
    We determine the potential parameters to reproduce the experimental spectra of
    charmonium.
    After that, we show theoretical results of the $\Lambda_c$ and $\Lambda_b$ baryons
    calculated by using the $q$-$Q$ type and $S$-$Q$ type potentials.
    We also calculate the excitation spectrum of the $\Xi_c$ baryon by using the $q$-$Q$ type potential.

    \subsection{Model parameters}
    \label{sec:modelparameter}
    The diquark inside a heavy baryon has the same color as the antiquark inside a quarkonium.
    Since the confinement force depends only on the color charge and not on the flavor in the first approximation,
    it is reasonable to use the same potential for the quark-diquark system as for the quark-antiquark system.
    \begin{table}[tbp]
        \centering
        \caption{Parameter sets of the potentials determined to reproduce
        the experimental spectrum of the charmonium.
        For parameter set 1, the non-confinement terms have
        a single parameter $\alpha_s$.
        Parameter set 2 introduces individual coupling constants for the non-confinement terms:
        $\alpha_{\rm Coul},~\alpha_{\rm Hyp},~\alpha_{\rm Dar},~
        \alpha_{\rm tens},~\alpha_{\rm LS}$ and $\alpha_{\rm oo}$.}
        \begin{tabular}{c|cccccc|c|c}
            \hline
            \hline
            &$\alpha_{\rm Coul}$ &$\alpha_{\rm Hyp}$ &$\alpha_{\rm Dar}$ &$\alpha_{\rm tens}$
            &$\alpha_{\rm LS}$ &$\alpha_{\rm oo}$ &$k$ [GeV/fm] &$\Lambda~[{\rm fm^{-1}}]$
            \\
            \hline
            No.1 &\multicolumn{6}{c|}{0.37} &0.9 &8.1\\
            No.2&0.40 &0.32 &0.22 &0.54&0.46 &0.65 &0.9 &6.4\\
            \hline
            \hline
        \end{tabular}  
        \label{tab:para_charm}
    \end{table}
    We determine the potential parameters to reproduce
    the experimental spectrum of charmonium.

    The parameters appearing in the potential
    are the string tension $k$ in the confinement part,
    the regularization parameter $\Lambda$,
    the fine structure constants $\alpha_s$ in the non-confinement part,
    and the masses of the heavy quark and the diquark.
    The string tension $k$ is fixed as $k=0.9~{\rm GeV/fm}$,
    which reproduces the global excitation spectra of the charmonium and bottonium \cite{Jido:2016yuv},
    and the charm quark mass is set to be $m_c=1.5~{\rm GeV}$.
    We consider two parameter sets. In parameter set 1 we assume a common coupling constant $\alpha_s$
    in the non-confining potential.
    We determine $\alpha_s$ and $\Lambda$ from the charmonium spectrum.
    In parameter set 2,
    we allow fine-tuning of the spectrum by introducing individual coupling constants
    for the non-confinement terms.
    The coupling constants, $\alpha_{\rm Coul},~\alpha_{\rm Dar},~\alpha_{\rm Hyp},~\alpha_{\rm LS},~\alpha_{\rm tems}$,
    and $\alpha_{\rm oo}$, are determined together with $\Lambda$ by the charmonium spectrum.
    The determined values are listed in Table \ref{tab:para_charm}.
    
\begin{figure*}[t]
        \centering
        \includegraphics[scale=0.6]{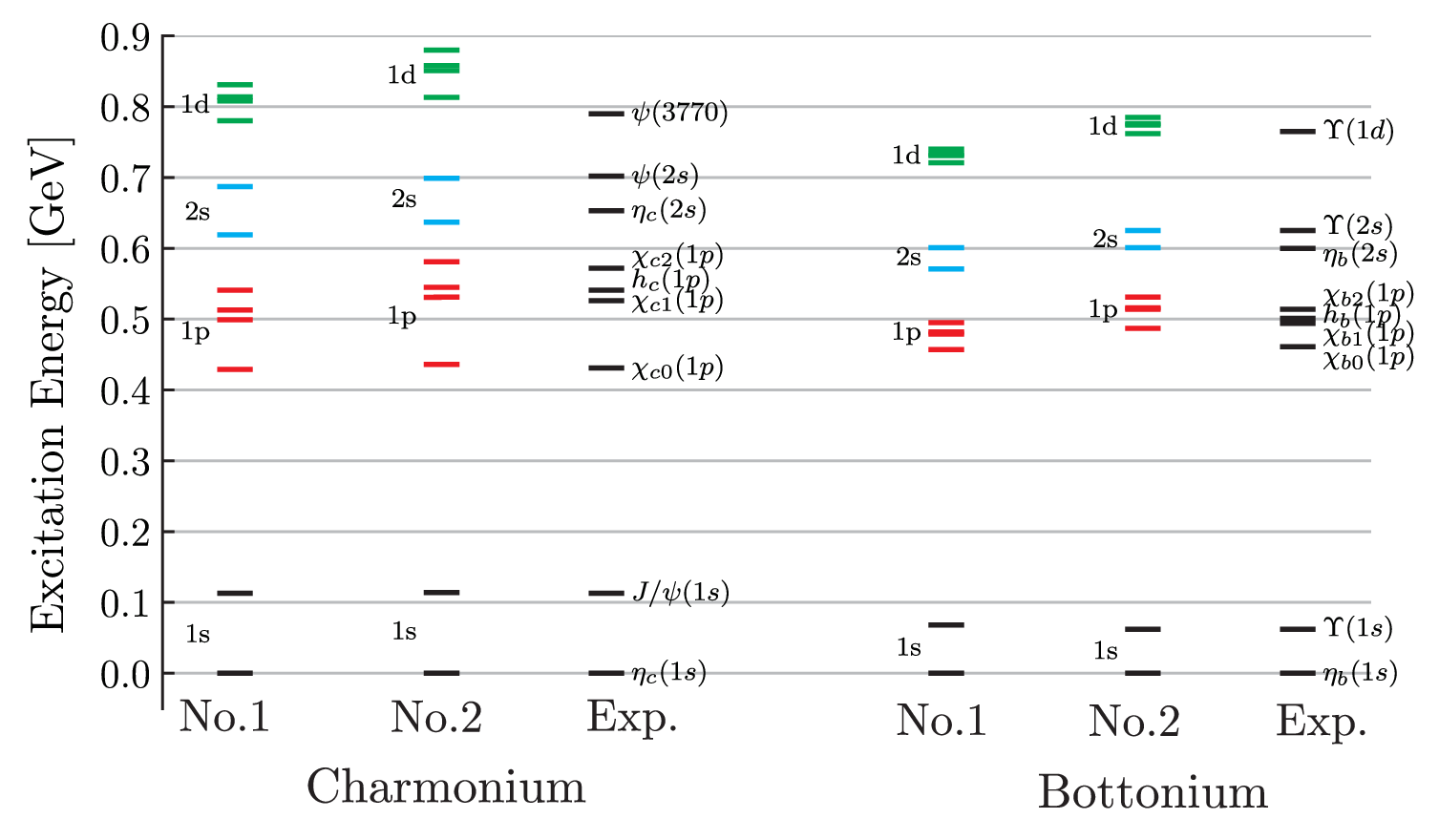}
        \caption{Excitation spectra of the charmonium and bottonium
        measured from the ground states.
        The masses of the charm bottom quarks are with $m_c=1.5$ GeV, and $m_b=4.0$ GeV, respectively.
        The parameters are determined to reproduce the experimental data
        of the charmonium.
        Two parameter sets are considered as shown in Table~\ref{tab:para_charm}.
        The experimental data are taken from the Particle Data Group~\cite{Workman:2022ynf}.}
        \label{fig:charm_bottom_spectra}
    \end{figure*}

    We show the
    excitation energy spectrum of charmonium calculated with the
    determined parameters in Fig.~\ref{fig:charm_bottom_spectra}, where
    the excitation energies are measured from the lowest state.
    We also show the excitation energy spectrum of bottonium obtained with the same parameter sets
    for comparison.
    The bottom quark mass is set as $m_b=4.0~{\rm GeV}$.
    As we can see, these parameters also reproduce the bottonium spectrum well.
    The spectra calculated with parameter set 2 reproduce the experimental values better
    than those with parameter set 1.

    \subsection{Excitation spectra of $\Lambda_c$ and $\Lambda_b$ with $q$-$Q$ type potential}
    \label{sec:spectra_heavybaryon_qq}
    In this subsection, we calculate the excitation spectra of the $\Lambda_c$ and $\Lambda_b$ baryons
    by using the $q$-$Q$ type potential.
    In this article, we assume that the mass of the non-strange scalar diquark is $m_d=0.5$ GeV.

\begin{figure*}[h]
        \centering
        \includegraphics[scale=0.5]{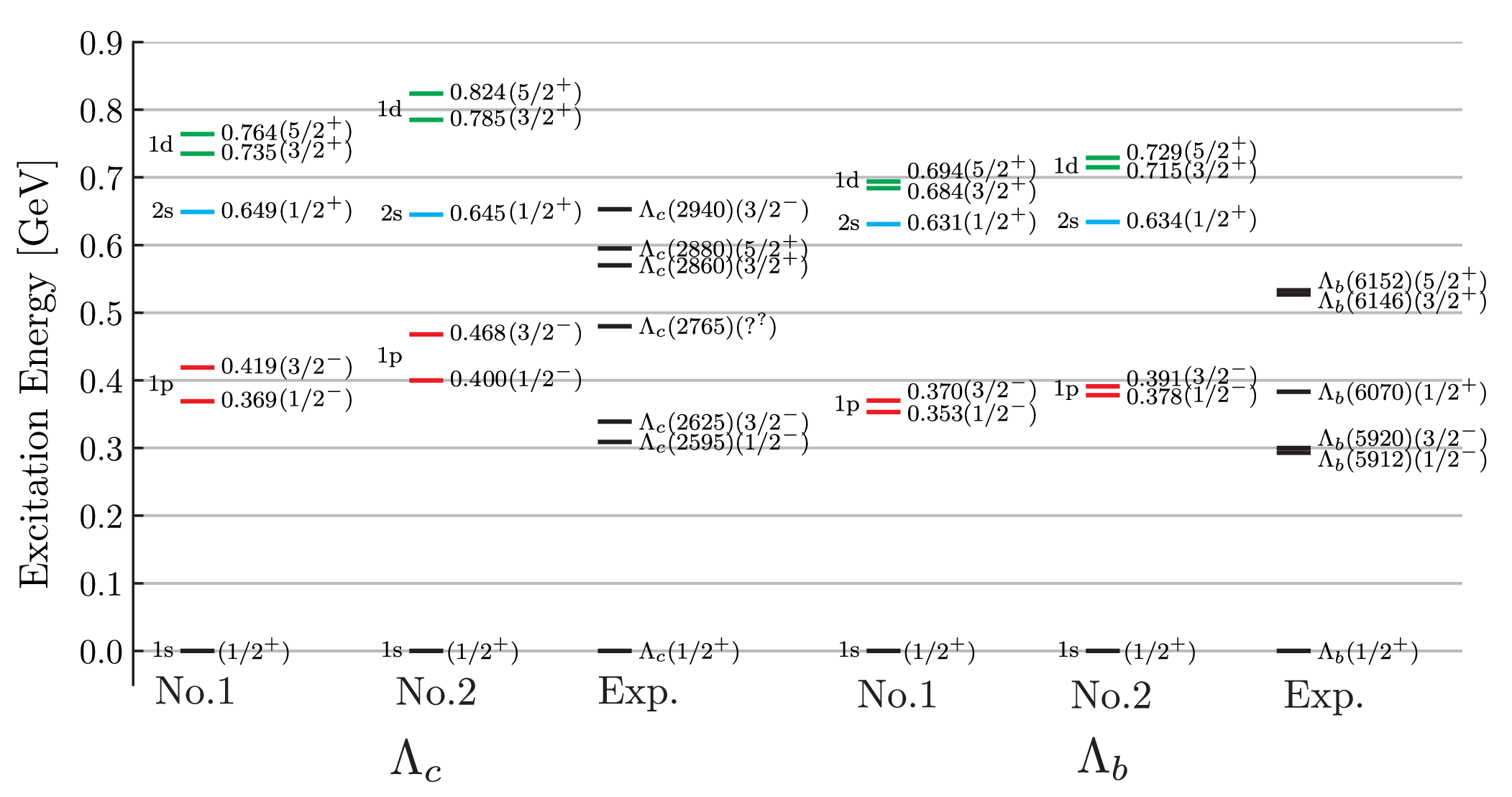}
        \caption{Calculated spectra with the $q$-$Q$ type potential for $\Lambda_c$ and $\Lambda_b$ systems.
        The parameter sets
        shown in Table~\ref{tab:para_charm} are used.
        The experimental data are taken
        from the Particle Data Group \cite{Workman:2022ynf}.
        }
        \label{fig:Lamc_Lamb_spectra}
    \end{figure*}

    Figure~\ref{fig:Lamc_Lamb_spectra}
    shows the excitation spectra of the $\Lambda_c$ and $\Lambda_b$ baryons
    calculated with the $q$-$Q$ type potential
    by using the parameter sets
    in Table~\ref{tab:para_charm}.
    As we can see,
    the $1p$ excitation energies and LS splitting are reproduced the experimental data
    better than the previous work~\cite{Jido:2016yuv}
    but they are slightly overestimated.
    The higher excitation energies are not reproduced.
    The $\Lambda_c~1p$ excitation energies for parameter set 1
    reproduce the experimental data better than those for parameter set 2,
    in contradiction to the results for the charmonium.
    This implies that it is hard to reproduce both spectra of the quarkonia
    and heavy hadrons by commom parameters.
    In Ref.~\cite{Jido:2016yuv}, the calculated excitation energy of the $\Lambda_c\ 1p$ state
    was overestimated to be about 140 MeV if one uses the parameters
    $\alpha_s=0.4$ and $k = 0.9~\rm{GeV/fm}$
    which is determined so as to reproduce the quarkonia spectra.
    In contrast, in our work, the energy difference between the experimental and
    theoretical values is about 70--115 MeV.
    The calculated $\Lambda_c\ 1p$ excitation energies are improved
    roughly 20\% or more by
    taking the relativistic corrections
    in the $q$-$Q$ type potential model.
    The relativistic correction is significant for the heavy baryon spectra.

\begin{figure*}[h]
        \centering
        \includegraphics[scale=0.5]{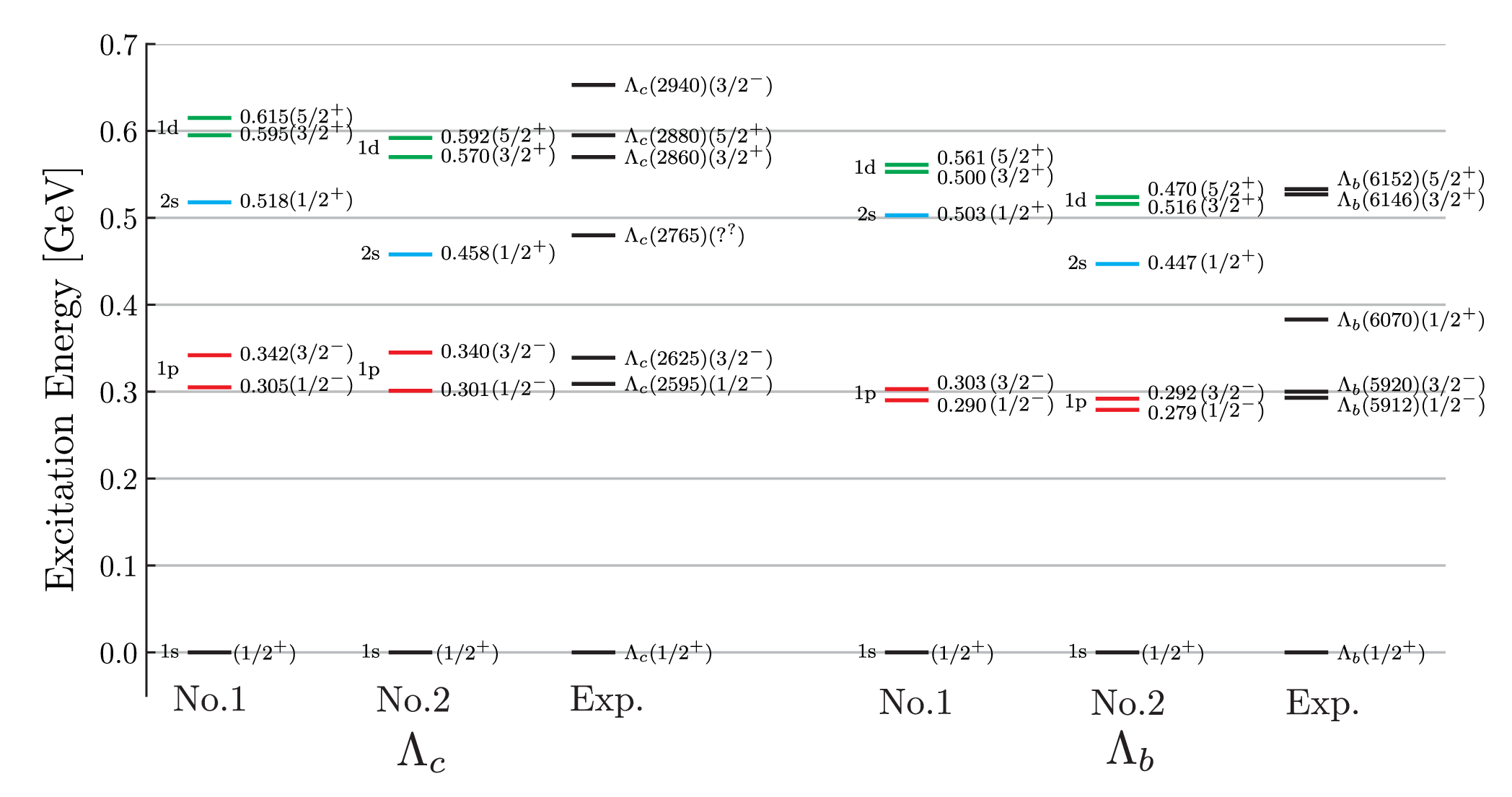}
        \caption{Calculated spectra with $q$-$Q$ type potential for
        the $\Lambda_c$ and $\Lambda_b$ baryon.
        The string tensions are redetermined to reproduce the $\Lambda_c~1p$ excitation energies
        for each parameter set.
        For parameter set 1, the value of the string tension is $k=0.63~\rm GeV/fm$, and
        for parameter set 2, that is $k=0.50~\rm GeV/fm$.
        Experimental data are taken from the Particle Data Group~\cite{Workman:2022ynf}.
        }
        \label{fig:Lamc_Lamb_spectraRekqq}
    \end{figure*}

    Next we discuss the LS splitting.
    We find that the LS splitting is slightly overestimated comparing the experimental data.
    We find that if we change the value of the parameter $\alpha_{\rm LS}$ of the LS term,
    the LS splitting of the $\Lambda_c$ baryon
    can be fitted simultaneously to the experimental data but $1p$ excitation energies cannot be done.
    Both of them can be fitted to the experimental data only by changing the strength of the string tension $k$,
    which is a similar situation with Ref.~\cite{Jido:2016yuv}.
    In Fig.~\ref{fig:Lamc_Lamb_spectraRekqq},
    we show the excitation spectra of
    the $\Lambda_c$ and $\Lambda_b$ baryons calculated with the $q$-$Q$ type potential
    by using parameters given in Table~\ref{tab:para_charm}
    but with different string tensions.
    The string tension is redetermined to reproduce the spin-weighted average
    of the excitation energies of the $\Lambda_c~1p$ states, 0.330 GeV.
    For parameter set 1, the string tension is found to be $k=0.63~\rm GeV/fm$, and
    for parameter set 2, it is to be $k=0.50~\rm GeV/fm$.
    As we can see, the LS splitting of the $\Lambda_c$ baryon matches the experimental data
    as well as the $1p$ excitation energies.
    The string tension have to be reduced by 30--40 \% to reproduce
    the experimental data of both $1p$ excitation energies and the LS splitting quite well,
    although its reduction is smaller than that invented in Ref.~\cite{Jido:2016yuv}.

    Here, let us move on the discussion for other excitation states.
    In contrast to the $1p$ excitation states, higher excitation states do not match the experimental data and
    they are overestimated considerably.
    Many studies have found that the theoretical spectra of $\Lambda_c(2765)$ are overestimated
    as the $\Lambda_c~2s$ state against the experimental observation,
    which is suggested to appear as a Roper-like state.
    Similarly, our result for the $\Lambda_c~2s$ state is about 1.4 times larger than
    the experimental data, which is consistent with those works.
    As for the $\Lambda_c(2880)$,
    Ref.~\cite{Nagahiro:2016nsx} evaluated
    the decay widths of charmed baryons from one-pion emission
    and found that the diquark inside $\Lambda_c(2880)$
    potentially having spin 1 can explain the small decay ratio between
    $\Lambda_c^*\rightarrow\Sigma_c^*(2520)\pi$ and
    $\Lambda_c^*\rightarrow\Sigma_c(2455)\pi$.
    This model also implies that the $\Lambda_c(2880)$ does not have the spin 0 diquark.

\begin{figure*}[tb]
        \centering
        \includegraphics[scale=0.6]{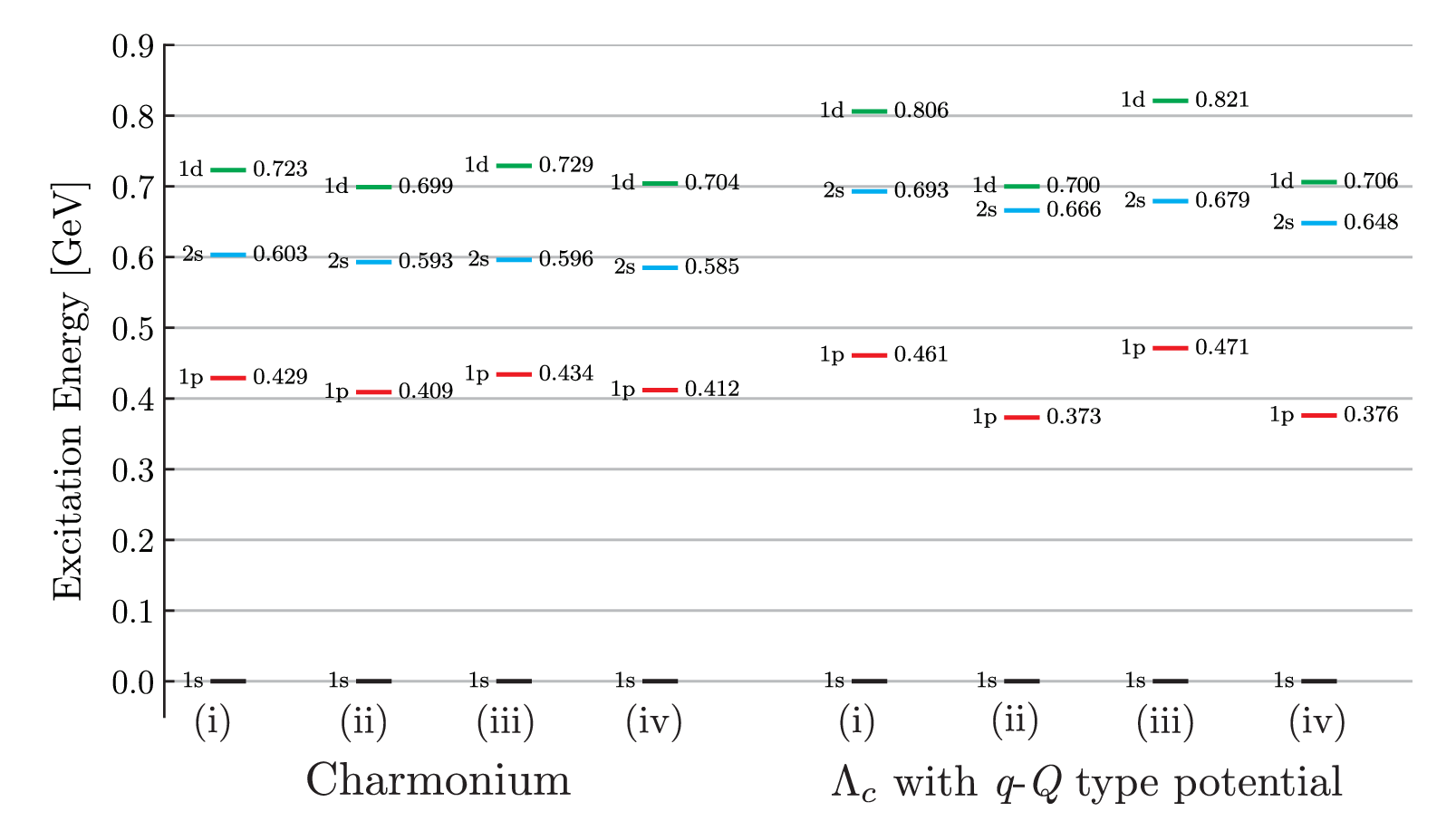}
        \caption{Excitation energies calculated with potentials (i)--(iv) listed in Eq.~(\ref{eq:pottype}).
        The charmonium spectra are on the left side, and the $\Lambda_c$ spectra are on the right side.
        The $\Lambda_c$ spectra are calculated with the $q$-$Q$ type potential.
        The potential parameters are fixed to $\alpha_s=0.4$, $k=0.9~\rm GeV/fm$, and
        $\Lambda=3.5~\rm fm^{-1}$.}
        \label{fig:Lamc_qq_potcomp}
    \end{figure*}

    Next we discuss the effect of each term of the potentials
    for the charmonium spectra and $\Lambda_c$ spectra.
    Figure~\ref{fig:Lamc_qq_potcomp} shows the excitation spectra of the charmonium
    and $\Lambda_c$ baryon
    calculated with the potentials as
    \begin{equation}
        \begin{split}
            &(\rm i)~V=V_{\rm conf}+V_{\rm Coul} \\
            &(\rm ii)~V=V_{\rm conf}+V_{\rm Coul}+V_{\rm Dar}\\
            &(\rm iii)~V=V_{\rm conf}+V_{\rm Coul}+V_{\rm oo}\\
            &(\rm iv)~V=V_{\rm conf}+V_{\rm Coul}+V_{\rm Dar}+V_{\rm oo}
        \end{split}
        \label{eq:pottype}
    \end{equation}
    by using the $q$-$Q$ type potential for the $\Lambda_c$ baryon.
    Here, we fix the parameters as $\alpha_s=0.4$, $k=0.9~\rm GeV/fm$,
    and $\Lambda=3.5~\rm fm^{-1}$.
    The values of parameters $\alpha_s$ and $k$ are used in Ref.~\cite{Jido:2016yuv} and
    the value of $\Lambda$
    is used in Ref.~\cite{Yoshida:2015tia}.
    The common parameter in the non-confinement part $\alpha_s$ is utilized
    to examine the effect of the potential term.
    We find that the effect of relativistic corrections
    for the charmonium spectra
    is rather small.
    From Fig.~\ref{fig:Lamc_qq_potcomp} (i) and (ii) on the charmonium part,
    we find that the level spacing between $1s$ and $1p$ becomes smaller
    owing to the Darwin term, which reduces the $1p$ excitation energy
    by 20 MeV.
    This is because the Darwin term is repulsive for $1s$ state and
    lifts its energy level up.
    Looking at (iii) on the charmonium part in Fig.~\ref{fig:Lamc_qq_potcomp},
    the orbit-orbit interaction increases the $1p$ excitation energy
    of the charmonium by about 5 MeV, which is very small.

    As for the $\Lambda_c$ part,
    the effect of the Darwin term is large for the $\Lambda_c$ spectra
    while that of the orbit-orbit interaction is small.
    The Darwin term reduces the $1p$ excitation energy
    by 88 MeV.
    The effect of the Darwin term for the $\Lambda_c$ excitation spectra is
    thus more significant than that for the charmonium
    because of the different mass dependence of the Darwin term as
    \begin{equation}
        V_{\rm Dar} = \frac{2\pi}{3}\alpha_{\rm Dar}
                \frac{2}{m_Q^2}
            \frac{2L+1}{4\pi r^2}\Lambda e^{-\Lambda^2 r^2}
    \end{equation}
    for the charmonium and
    \begin{equation}
        V_{\rm Dar} = \frac{2\pi}{3}\alpha_{\rm Dar}\left(
                \frac{1}{m_Q^2}+\frac{4}{m_d^2}
            \right)
            \frac{2L+1}{4\pi r^2}\Lambda e^{-\Lambda^2 r^2}
    \end{equation}
    for the $\Lambda_c$ baryon.
    We can see that the Darwin term for the $\Lambda_c$ baryon contains
    the term depending on the diquark mass ${1}/{m^2_d}$
    which remains finite even in the heavy quark limit ($m_h\rightarrow\infty$).
    This term provides large influence for the $\Lambda_c$ spectra.
    The orbit-orbit interaction whose effect on the $\Lambda_c~1p$ excitation spectra is 10 MeV,
    has small effect on the excitation spectra of the $\Lambda_c$ baryon
    with the $q$-$Q$ type potential.
    Thus, the Darwin term is not so important for the charmonium, but important for the $\Lambda_c$.

    In conclusion, the relativistic corrections are not enough large to reproduce
    the $\Lambda_c$ $1p$ excitation energy, but thanks to the effect of the Darwin term,
    it is unnecessary to reduce the string tension
    in a quark-diquark system to as much as
    half of that in the quark-antiquark system
    to reproduce the $1p$ excitation energy of $\Lambda_c$ baryon.
    The existence of the Darwin term related to the diquark mass
    can be one of the keys of the solution to the string tension puzzle.
    

    \subsection{Excitation spectra of $\Lambda_c$ and $\Lambda_b$ with $S$-$Q$ type potential}
    \label{sec:spectra_heavybaryon_Sq}
    In this subsection, we discuss the energy spectra of the $\Lambda_c$ and $\Lambda_b$ baryons
    calculated by using the $S$-$Q$ type potential.
        
\begin{figure*}[h]
        \centering
        \includegraphics[scale=0.55]{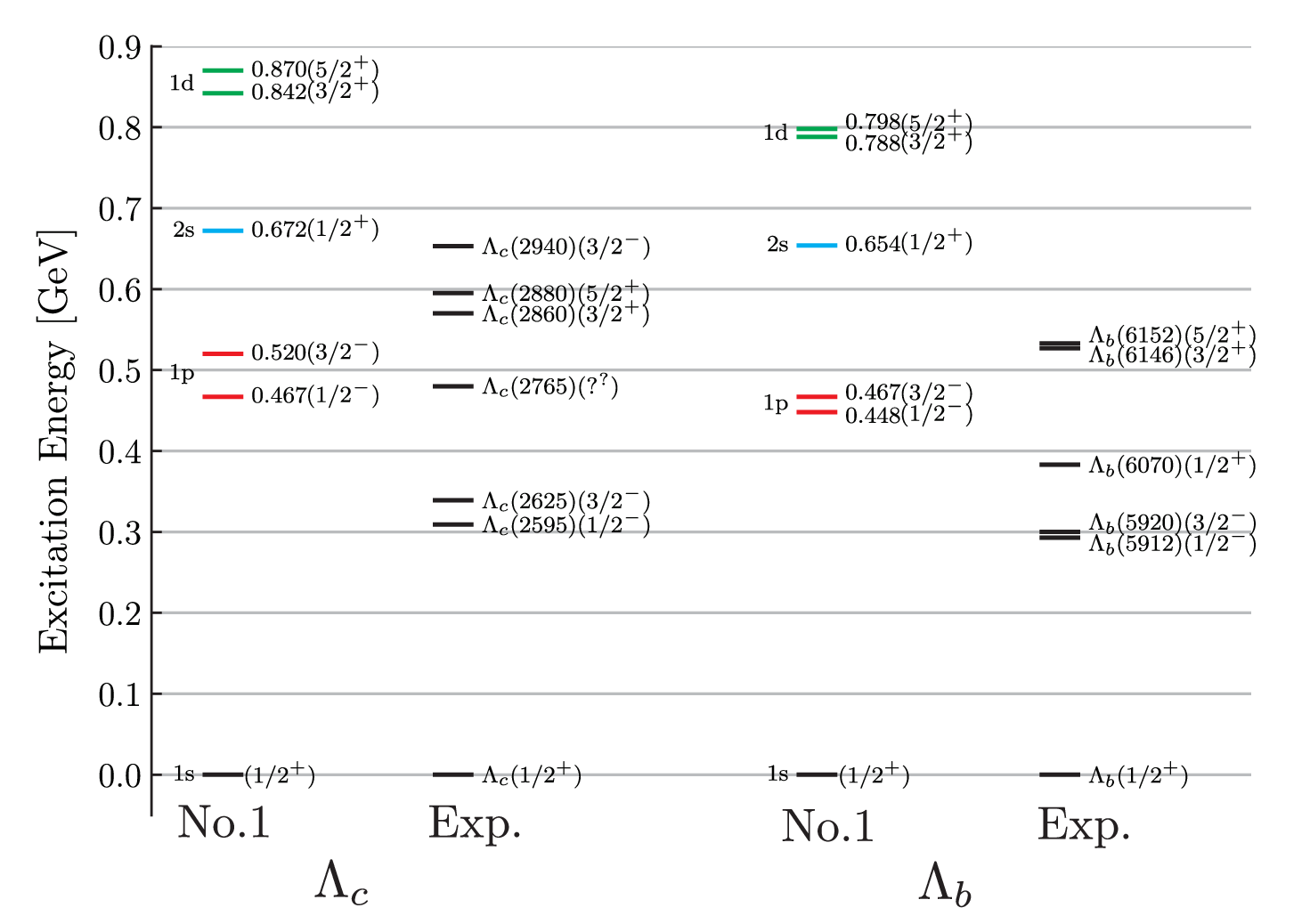}
        \caption{Calculated spectra with the $S$-$Q$ type potential for
        $\Lambda_c$ and $\Lambda_b$ systems.
        The experimental data are taken
        from the Particle Data Group~\cite{Workman:2022ynf}.
        Parameter set 1
        shown in Table~\ref{tab:para_charm} is used.
        }
        \label{fig:Lamc_Lamb_Sqspectra}
    \end{figure*}

    Figure~\ref{fig:Lamc_Lamb_Sqspectra} shows the
    $\Lambda_c$ and $\Lambda_b$ excitation spectra calculated with the $S$-$Q$ type potential
    for parameter set 1.
    As we can see, all the calculated excitation energies are larger than the experimental values.
    Focusing on the $\Lambda_c\ 1p$ excited energies, we find that
    the energy difference between the calculated results and the experimental data
    is about 170 MeV, which is 1.2 times as large as that in Ref.~\cite{Jido:2016yuv}.

\begin{figure}
        \centering
        \includegraphics[scale=0.5]{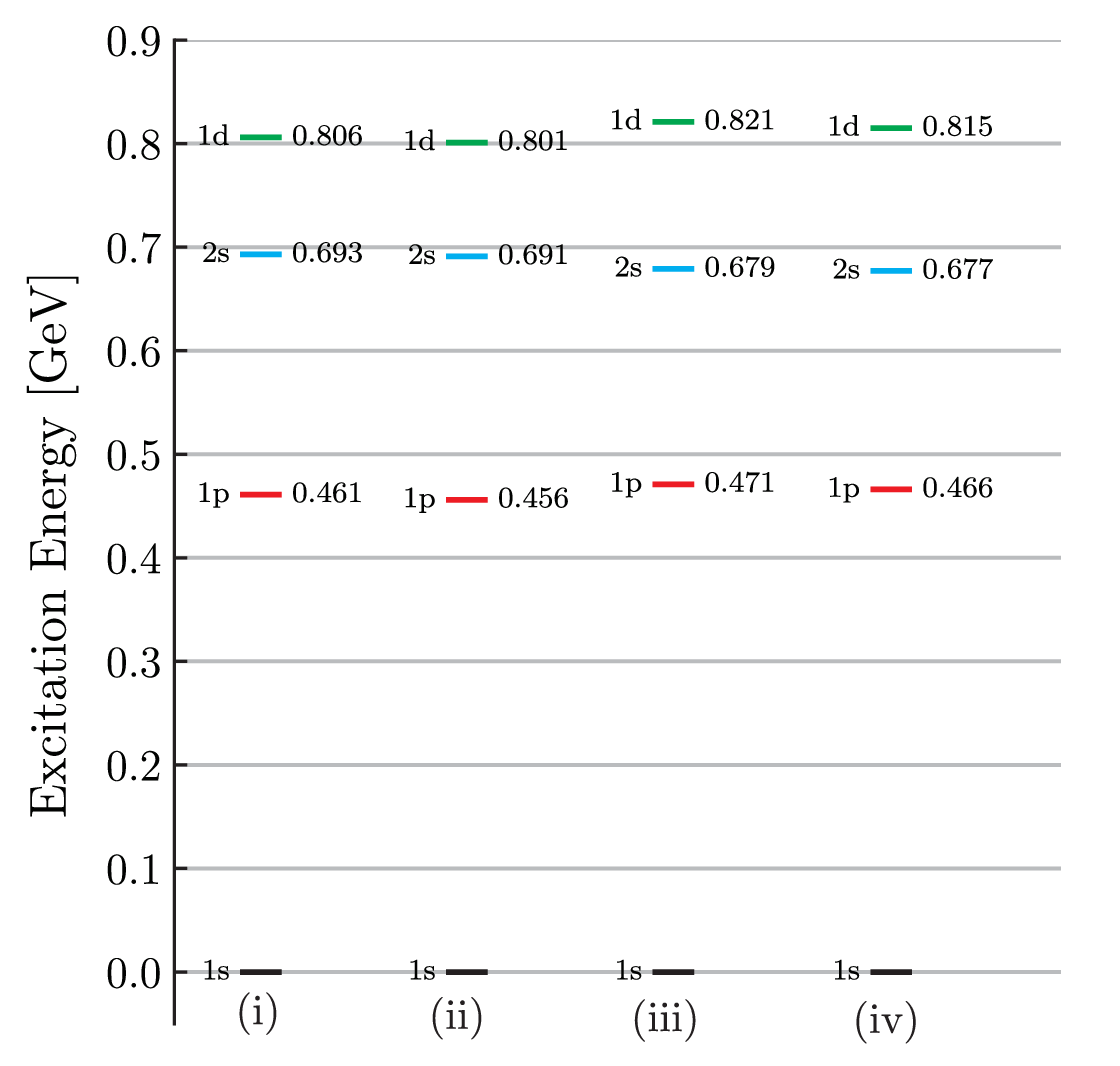}
        \caption{Excitation energies of $\Lambda_c$ calculated with potentials (i)--(iv) in Eq.~(\ref{eq:pottype})
        for the $S$-$Q$ type potential.
        The potential parameters are fixed to $\alpha_s=0.4$, $k=0.9~\rm GeV/fm$,
        and $\Lambda= 3.5~\rm fm^{-1}$.
        }
        \label{fig:Lamc_Sq_potcomp}
    \end{figure}

    Next we discuss each contribution of the relativistic correction.
    Figure~\ref{fig:Lamc_Sq_potcomp} shows the excitation spectra of $\Lambda_c$ baryon
    calculated by using the potentials (i)--(iv) listed in Eq.~(\ref{eq:pottype})
    whose terms are taken from the $S$-$Q$ type potential, and
    we use parameters $\alpha_s=0.4$, $k=0.9~{\rm GeV/fm}$,
    and $\Lambda=3.5~{\rm fm^{-1}}$
    to see effects of each term for the excitation spectra of $\Lambda_c$.
    Effects of all terms of the potential are small, and they are 5 MeV.
    Comparing (ii) on the $\Lambda_c$ part in Fig.~\ref{fig:Lamc_qq_potcomp}
    to that in Fig.~\ref{fig:Lamc_Sq_potcomp},
    in contrast to the $q$-$Q$ type potential,
    the effect of the Darwin term on the $S$-$Q$ type potential is much smaller.
    The term depending on the diquark mass in the Darwin term does not
    exist in the $S$-$Q$ type potential (\ref{eq:Derwin_Sq}), and
    this causes a large difference between the heavy baryon spectra
    with the $q$-$Q$ and $S$-$Q$ type potentials.
    Table~\ref{tab:ExpectationValue_oo} shows the expectation values of the orbit-orbit interaction
    for each state of the charmonium and the $\Lambda_c$ baryon
    with the $S$-$Q$ type potential.
    As we can see, since the orbit-orbit interaction is attractive and
    its effects for the $s$ state are a bit larger,
    the energy differences between the $1s$ state and the $1p$ and $1d$ states
    become enhanced.
    Thus,
    the $\Lambda_c~1p$ excitation energies calculated with the $S$-$Q$ type potential
    are larger than those calculated in Ref.~\cite{Jido:2016yuv}.
    \begin{table}[tb]
        \centering
        \caption{Expectation values of the orbit-orbit interaction for each state of the charmonium
        and the $\Lambda_c$ baryon calculated with $S$-$Q$ type potential in unit of GeV.
        Parameters $\alpha_s=0.4$, $k=0.9~{\rm GeV/fm}$,
        and $\Lambda=3.5~{\rm fm^{-1}}$ are used.
        }
        \begin{tabular}{c|cccc}
            \hline
            \hline
            &$1s$ &$1p$ &$2s$ &$1d$ \\
            \hline
            $c\bar{c}$ &-0.030 &-0.026 &-0.036 &-0.024 \\
            $\Lambda_c$ with $S$-$Q$ type &-0.051 &-0.041 &-0.064 &-0.037 \\
            \hline
            \hline
        \end{tabular}
        \label{tab:ExpectationValue_oo}
    \end{table}
    Nevertheless, we find that the magnitudes of
    the relativistic corrections for the $S$-$Q$ type potential
    is not so large.

\begin{figure*}[htbp]
        \includegraphics[scale=0.55]{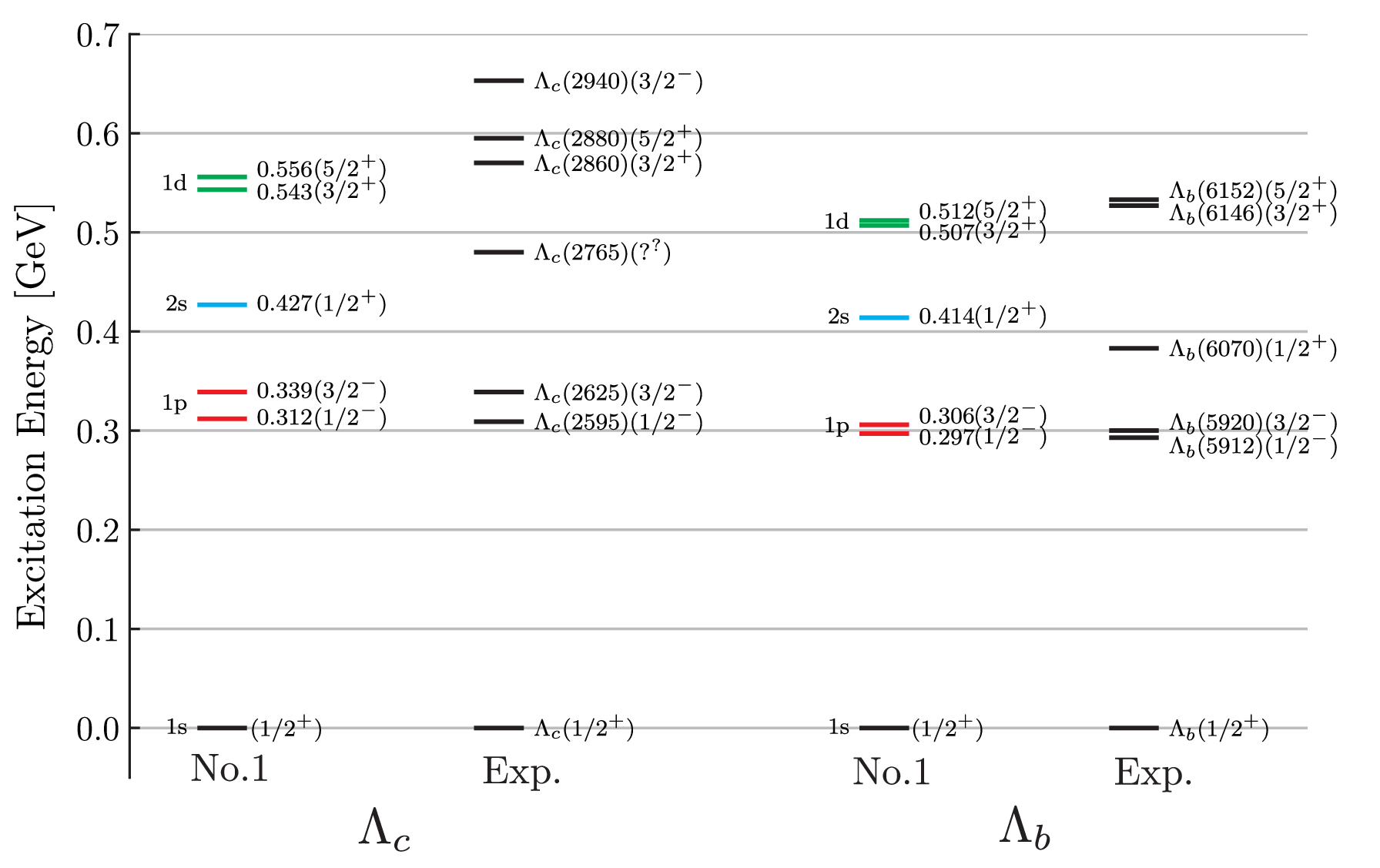}
        \caption{Calculated spectra with the $S$-$q$ type potential for
        the $\Lambda_c$ and $\Lambda_b$ baryons.
        The
        string tension is redetermined to reproduce the $\Lambda_c~1p$ excitation energies.
        The value of the string tension is $k=0.42~\rm GeV/fm$.
        Experimental data are taken from the Particle Data Group~\cite{Workman:2022ynf}.
        }
        \label{fig:Sqspectra_Rek}
    \end{figure*}

    As done in Ref.~\cite{Jido:2016yuv}, we redetermine the string tension to reproduce the spin-weighted average
    of the excitation energies of the $\Lambda_c~1p$ states, 0.330 GeV.
    To reproduce the experimental data, we find
    the string tension $k$ is needed to become smaller $k=0.42$ $\rm GeV/fm$ for parameter set 1.
    Figure~\ref{fig:Sqspectra_Rek} shows the calculated excitation spectra with
    this string tension.

    In conclusion, the relativistic corrections for the $S$-$Q$ type potential are too small
    to solve the string tension puzzle, so
    we have to reduce the string tension $k$ to reproduce the experimental spectrum with
    the $S$-$Q$ type potential.
    And then, the string tension puzzle still remains.
    We can see that there is a large difference between the two potential models stemming from
    treating the color structure of the diquark.

    \subsection{Excitation spectra of $\Xi_c$ baryon}
    \label{sec:Xic}
    In this subsection, we show the calculated result of the $\Xi_c$ excitation spectrum
    with the $q$-$Q$ type potential.
    We consider that the $\Xi_c$ baryon consists of a charm quark and a scalar strange diquark
    which is composed of a light quark and a strange quark.

    We estimate the strange diquark mass $m_{ds}$
    by the difference of the ground state
    mass of $\Lambda_c$ and the isospin averaged mass of $\Xi_c$, which is 0.18 GeV.
    The ground state mass of the $\Lambda_c$ baryon is calculated
    by using the value of the diquark mass 0.5 GeV and parameter set 1.
    We determine the strange diquark mass by fitting the mass difference
    between the ground states of the $\Xi_c$ and $\Lambda_c$ baryons to the experimental value,
    and we find the strange diquark mass to be 0.81 GeV for parameter set 1.

\begin{figure}[h]
        \centering
        \includegraphics[scale=0.45]{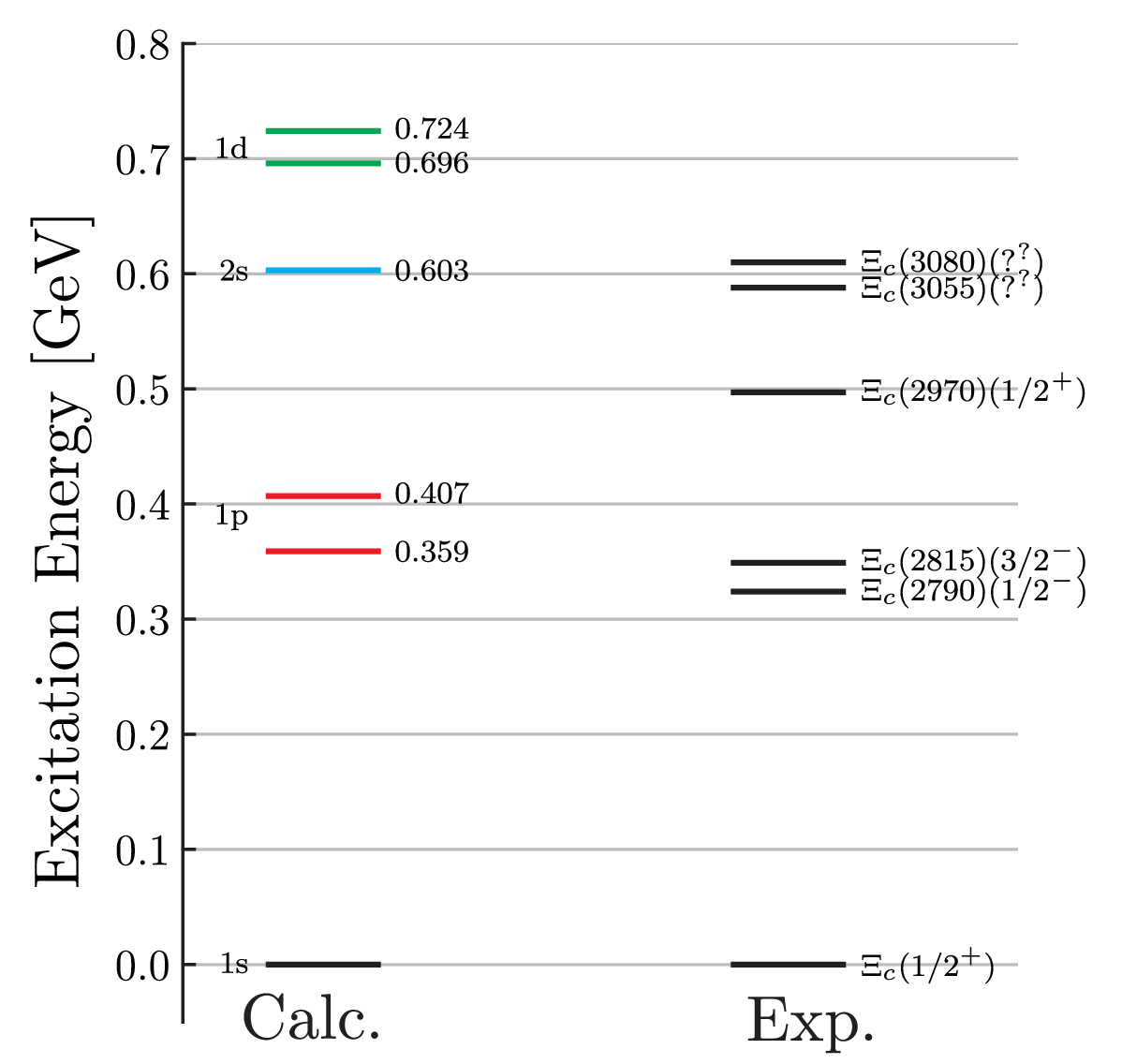}
        \caption{Calculated excitation spectra with the $q$-$Q$ type potential for the $\Xi_c$ baryon.
        The $q$-$Q$ type potential with
        parameter set 1 as in Table.~\ref{tab:para_charm} is used to calculate them.
        The experimental data are taken
        from the Particle Data Group~\cite{Workman:2022ynf}.
        }
        \label{fig:Xic_spectrum}
    \end{figure}

    Figure~\ref{fig:Xic_spectrum} shows the excitation spectrum of the $\Xi_c$ baryon calculated
    with parameter set 1 listed in Table~\ref{tab:para_charm}.
    We find that the calculation a bit overestimates the $1p$ excitation energies and their difference
    between the calculation and the experiments is 45 MeV.
    Even though the average of the calculated $1p$ excitation energies is consistent with the experimental one,
    the LS splitting of the $1p$ states in the calculation is about twice larger that that of the experiments.
    Similarly to the result of the $\Lambda_c$ excitation spectra with the $q$-$Q$ type potential,
    in order to reproduce the $1p$ excitation energies and the LS splitting,
    we should reduce the string tension.
    For the higher excited states of the $\Xi_c$ baryon,
    the calculation overestimates largely the excitation spectra.
    Similarly to the result of $\Lambda_c$ excitation spectra with the $q$-$Q$ type potential,
    if both of calculated results of the $1p$ excitation energies and the LS splitting reproduce
    experimental data, we should reduce the string tension.
    For the higher excitation states of the $\Xi_c$ baryon,
    the calculated values are larger than the experimental data.

    \section{summary}
    \label{sec:summary}
    In this paper, we have calculated the excitation spectra of the heavy baryons
    in a quark-diquark model with relativistic corrections.
    The heavy baryon has been assumed to be composed of a heavy quark and a point-like scalar diquark
    having anti-color $\bar{\bf 3}$ and spin $S=0$.
    We considered two types of relativistic corrections by
    depending on
    the internal color structure of the diquark.
    In the first approach, the diquark
    has been treated as an exact two-quark pair,
    and this model is called as $q$-$Q$ type potential.
    In the second approach, the diquark has been treated as just a scalar particle having
    no internal structure, and
    this model is called as $S$-$Q$ type potential.
    Our objective is to obtain a solution to the puzzle pointed out in Ref.~\cite{Jido:2016yuv},
    which we call as the string tension puzzle,
    by considering relativistic corrections for the quark-diquark potential.
    
    For the $q$-$Q$ type potential, we have found that
    the calculated $\Lambda_c~1p$ excitation energies
    are slightly larger than the experimental data.
    We need to reduce the string tension to reproduce the experimental data
    of both $1p$ excitation energies and the LS splitting with good precision,
    but we do not need to reduce it by half of the strength in the quark-antiquark system.
    This potential has a Darwin term with the diquark mass.
    This originates from the quark structure of the diquark,
    which indicates that the Darwin effect is stronger for heavy baryons than for quarkonia.
    The energy difference between the $\Lambda_c~1s$ state and the $1p$ state
    becomes relatively small,
    as the energy of the $1s$ state increases due to the Darwin term.
    This demonstrates that considering the relativistic effects, especially the Darwin term,
    is important for solving the string tension puzzle.

    The $S$-$Q$ type potential in which the diquark has been treated as a
    scalar particle does not have a Darwin term
    with the diquark mass.
    This means that the string tension on the $S$-$Q$ type potential should be taken
    much smaller than one in the quark-antiquark potential, which
    is qualitatively the same result as the previous work~\cite{Jido:2016yuv}.
    Our findings have shown that treating the internal color structure of the diquark causes large differences,
    and since the diquark is composed of two quarks,
    its structure should be carefully considered.
    
    We have also calculated the $\Xi_c$ excitation spectra by using the $q$-$Q$ type potential.
    The $\Xi_c$ baryon was considered as the bound system of the scalar strange diquark and the charm quark.
    We have found that the calculated $1p$ excitation energies
    reproduced the experimental data and the LS splitting was overestimated.
    The consistent results with the $\Lambda_c$ baryon has been obtained.


\begin{acknowledgments}
    This work was supported by
    JST, the establishment of university fellowships towards the creation 
of science technology innovation, Grant No. JPMJFS2127.
the work of H. N. was supported by the Grants-in-Aid for Scientific Research [Grant No. JP21K03536],
and the work of D. J. was partially supported by Grants-in-Aid for Scientific Research [No. JP21K03530, JP22H04917].
\end{acknowledgments}
\nocite{*}

\bibliography{Diquark_quark}

\end{document}